\documentclass[preprint,preprintnumbers,amsmath,amssymb,nofootinbib]{revtex4}
\pdfoutput=1
\usepackage{graphicx}
\usepackage[space]{grffile}

\usepackage{amsmath,amssymb}
\usepackage{mathptmx}
\usepackage[EULERGREEK]{sansmath}

\begin{document}

\begin{center}
\textbf{\Large Realizing a lattice spin model with polar molecules}\\*
\vspace*{2mm}
{\small Bo Yan,$^{1}$ Steven A. Moses,$^{1}$ Bryce Gadway,$^{1}$ Jacob P. Covey,$^{1}$ Kaden R. A. Hazzard,$^{1}$ Ana Maria Rey,$^{1}$ Deborah S. Jin,$^{1}$ Jun Ye$^{1}$}

\vspace{2mm}
\textit{\scriptsize $^1$JILA, National Institute of Standards and Technology and University of Colorado, and Department of Physics, University of Colorado, Boulder, Colorado 80309, USA.}\\*
\end{center}

\vspace {5 mm}

\begin{abstract}
\textbf{
With the recent production of polar molecules in the quantum regime~\cite{ni2008,silke2010}, long-range dipolar interactions are expected to facilitate the understanding of strongly interacting many-body quantum systems and to realize lattice spin models~\cite{Micheli_NP_2006} for exploring quantum magnetism. In atomic systems, where interactions require wave function overlap, effective spin interactions on a lattice can be realized through superexchange; however, the coupling is relatively weak and limited to nearest-neighbor interactions~\cite{trotzky:time-resolved_2008,fukuhara:quantum_2013,greif:quantum_2012}.  In contrast, dipolar interactions exist even in the absence of tunneling and extend beyond nearest neighbors. This allows coherent spin dynamics to persist even for gases with relatively high entropy and low lattice filling.  While measured effects of dipolar interactions in ultracold molecular gases have thus far been limited to the modification of inelastic collisions and chemical reactions~\cite{ni2010,miranda2011}, we now report the first observation of dipolar interactions of polar molecules pinned in a three-dimensional optical lattice.  We realize a lattice spin model where spin is encoded in rotational states of molecules that are prepared and probed by microwaves. This interaction arises from the resonant exchange of rotational angular momentum between two molecules and realizes a spin-exchange interaction. The dipolar interactions are apparent in the evolution of the spin coherence, where we observe clear oscillations in addition to an overall decay of the coherence.  The frequency of these oscillations, the strong dependence of the spin coherence time on the lattice filling factor, and the effect of a multi-pulse sequence designed to reverse dynamics due to two-body exchange interactions all provide clear evidence of dipolar interactions. Furthermore, we demonstrate the suppression of loss in weak lattices due to a quantum Zeno mechanism~\cite{misra:zeno_1988}. Measurements of these tunneling-induced losses allow us to independently determine the lattice filling factor.
The results reported here comprise an initial exploration of the behavior of many-body spin models with direct, long-range spin interactions and lay the groundwork for future studies of many-body dynamics in spin lattices.}
\end{abstract}

\maketitle

Long-range and spatially anisotropic dipole-dipole interactions permit new approaches for the preparation and exploration of strongly correlated quantum matter that exhibit intriguing phenomena such as quantum magnetism, exotic superfluidity, and topological phases~\cite{carr:cold_2009,lahaye_physics_2009,potter:2006,tjmodel,topological,
levinsen:topological_2011,knap:clustered_2012}. Ultracold gases of polar molecules provide highly controllable, long-lived, and strongly interacting dipolar systems and have recently attracted intense scientific interest.  Samples of fermionic $^{40}$K$^{87}$Rb polar molecules, with an electric dipole moment of 0.57~Debye (1~Debye $=3.336\times10^{-30}$ C$\cdot$ m)~\cite{ni2008}, have been prepared near the Fermi temperature, and all of the degrees of freedom (electronic, vibrational, rotational, hyperfine, and external motion) can be controlled at the level of single quantum states~\cite{hyperfine,miranda2011,chotia:long-lived_2012}.

The surprising discovery of bi-molecular chemical reactions of KRb at ultralow temperatures~\cite{silke2010,ni2010,miranda2011} appeared to be a major challenge for creating novel quantum matter. However, the molecules' motion, and consequently their reactions, can be fully suppressed in a three-dimensional (3D) optical lattice, where relatively long lifetimes ($>$25~s) have been observed~\cite{chotia:long-lived_2012}. The long-range dipolar interaction can then play the dominant role in the dynamics of the molecular internal degrees of freedom, for example, by exchanging two neighboring molecules' rotational states. With spin encoded in the rotational states of the molecule, these dipolar interactions give rise to spin-exchange interactions, analogous to those that play an important role in quantum magnetism and high-temperature superconductivity~\cite{superconductivity}.  In a 3D lattice, where each molecule is surrounded by many neighboring sites, this system represents an intriguing many-body quantum spin system where excitations can exhibit strong correlations even at substantially less than unit lattice filling~\cite{hazzard:far-from-equilibrium_2013}.

Several features distinguish the interactions in a molecular spin model from those observed in ultracold atomic systems.  For the superexchange interaction of atoms in optical lattices~\cite{trotzky:time-resolved_2008,fukuhara:quantum_2013,greif:quantum_2012}, the short-range nature of the interparticle interactions necessitates a second-order perturbative process in the tunneling of atoms between lattice sites. Hence, the energy scale of the superexchange interaction decreases exponentially with lattice depth. This spin-motion coupling limits superexchange to nearest-neighbor interactions and necessitates extremely low temperature and entropy.

In contrast, long-range dipolar interactions decay as $1/r^3$ with separation $r$, and interactions beyond nearest neighbors are significant. This long-range interaction allows exploration of coherent spin dynamics in very deep lattices where the molecules' translational motion is frozen and where the absence of tunneling would preclude the superexchange interactions of atoms. We note that the dipolar interaction is also unique compared to that of electrons, where an effective spin interaction arises due to the spin-independent Coulomb interaction and the exchange symmetry of the fermionic electrons.  In contrast, the dipolar interaction is a direct spin-spin interaction that does not require any wave function overlap. In addition to polar molecules, ultracold systems such as Rydberg atoms~\cite{saffman:quantum_rydberg_2010}, magnetic atoms~\cite{lahaye_physics_2009,lu:quantum_Dy_2012,
aikawa:bose-einstein_Er_2012,paz:resonant_2013}, and trapped ions~\cite{lanyon:universal_ion_2011,britton:engineered_ions_2012,islam:emergence_2013}, are candidates for realizing coherent, controllable spin models with power-law interactions; however, spin-exchange interactions have yet to be created and observed in these systems.

The molecular rotational states $|N,m_N\rangle$, where $N$ is the principal quantum number and $m_N$ the projection onto the quantization axis, are the focus of our current investigation of a dipolar spin system. In general, an external DC electric field induces a dipole moment in the laboratory frame by mixing opposite-parity rotational states.  However, even in the absence of a DC electric field, dipolar interactions can be established using a microwave field to create a coherent superposition between two rotational states, labeled as $|\uparrow \rangle$ and $|\downarrow \rangle$~\cite{barnett:quantum_2006}. In addition, a microwave field can probe the coherent spin dynamics due to dipolar interactions.

\begin{figure}
\includegraphics[height=12cm]{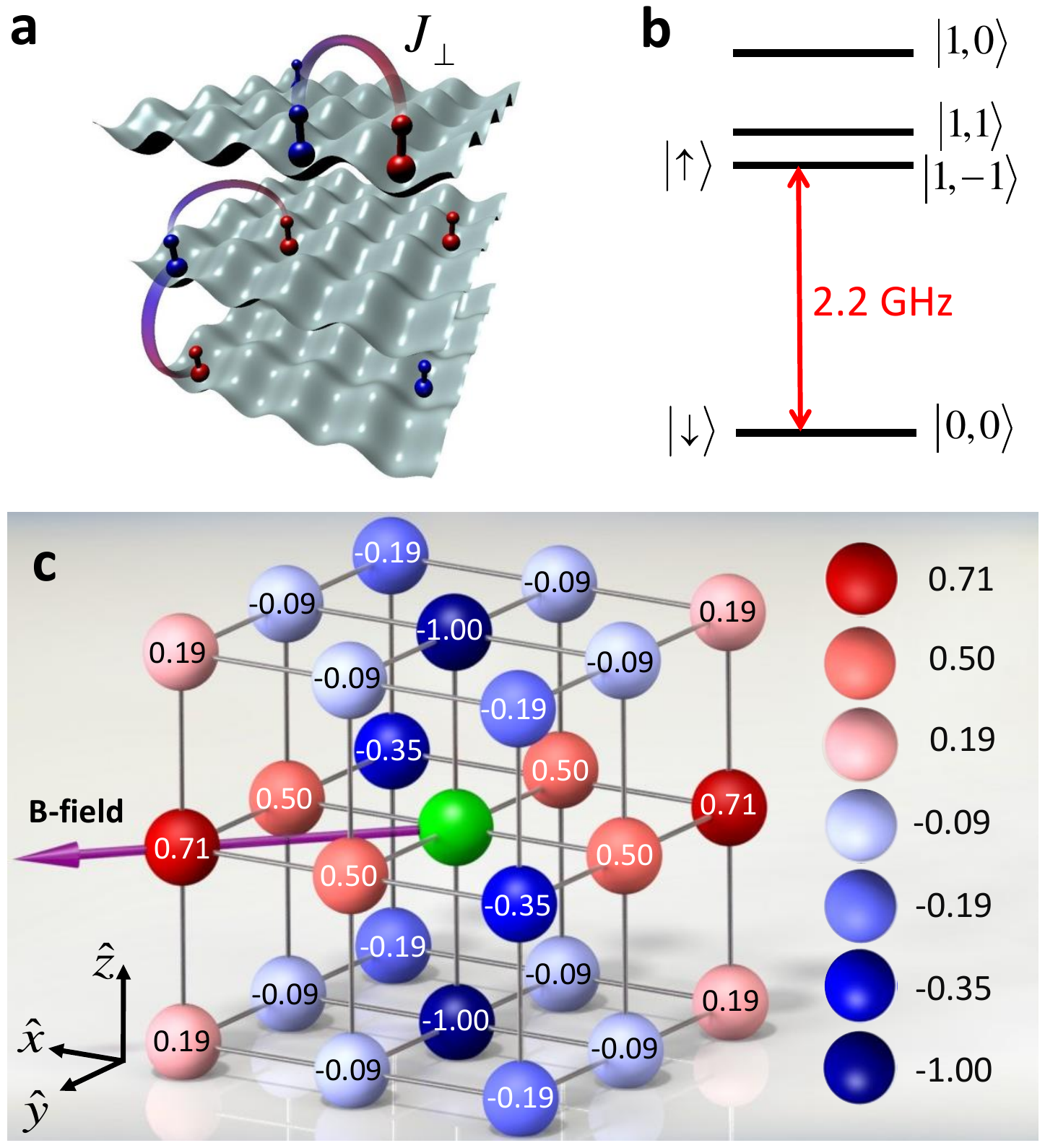}
\caption{\label{fig:scheme}
\textbf{Dipolar interactions of polar molecules in a 3D lattice.}
\textbf{~a},
Polar molecules are loaded into a deep 3D optical lattice. Microwaves are used to address the transition between two rotational states (red and blue represent different rotational states). $J_\perp$ characterizes the spin-exchange interaction energy. \textbf{b}, A schematic energy diagram (not to scale) is shown for the ground and first-excited rotational states. The degeneracy of the excited rotational states is broken due to a weak coupling of the nuclear and rotational degrees of freedom.
We use $|0,0\rangle$ and $|1,-1\rangle$ as our two spin states.
\textbf{c}, The interaction energy between any two molecules depends on their relative position in the lattice. The numbers shown give the geometrical factor $-V_{dd}(\mathbf{r}_i-\mathbf{r}_j)$ for the dipolar interaction of each site relative to the central site (green), under the specific quantization axis (B-field). Negative values (blue) correspond to attractive interactions, and positive values (red) to repulsive interactions.}
\end{figure}

In the absence of an applied electric field, two-level polar molecules trapped in a strong 3D lattice (Fig.~\ref{fig:scheme}a) can be described as a spin-1/2 lattice model
with the interaction Hamiltonian~\cite{barnett:quantum_2006,tjmodel,hazzard:far-from-equilibrium_2013}:
\begin{equation}\label{eqn:hamitonian}
H=\frac{J_\perp}{2}\sum_{i> j}V_{dd}(\mathbf{r}_i-\mathbf{r}_j)\left(S_i^+S_j^-+S_i^-S_j^+\right),
\end{equation}
where $S_i^{\pm}$ (along with $S_i^z$) are the usual spin-1/2 angular momentum operators on site $i$.  The dipolar interaction energy includes a geometrical factor $V_{dd}(\mathbf{r}_i-\mathbf{r}_j)=\left(1-3\cos^2\Theta_{ij}\right)/ |{\mathbf{r}_i-\mathbf{r}_j}|^3$, where $\mathbf{r}_i$ is the position of the $i^{th}$ molecule in units of the lattice constant $a$ and $\Theta_{ij}$ is the angle between the quantization axis $\hat{z}$ and the vector connecting molecules $i$ and $j$.
Generic spin-1/2 models also include a $J_z S_i^zS_j^z$ term; however, for polar molecules in the absence of an electric field, $J_z=0$ and the Hamiltonian reduces to the limiting case known as the spin-1/2 quantum XY model.  Here, the spin-exchange interaction is characterized by $J_{\perp}=-d_{\downarrow\uparrow}^2/4\pi\epsilon_0 a^3$,  where $\epsilon_0$ is the permittivity of free space and $d_{\downarrow\uparrow}=\langle\downarrow|d|\uparrow\rangle$ is the dipole matrix element between $|\downarrow\rangle$ and $|\uparrow\rangle$. Physically, this term is responsible for exchanging the spins of two trapped molecules (Fig.~\ref{fig:scheme}a).

In our experiment, we create $2\times 10^4$ ground-state KRb molecules in the lowest motional band of a 3D lattice formed by three mutually orthogonal standing waves at $\lambda = 1064$~nm. The lattice constant is $a=\lambda/2$ and the lattice depth is $40 \ E_r$ in each direction,  where $E_r=\hbar^2 k^2/2m$ is the recoil energy, $\hbar$ is the reduced Planck constant, $k=2 \pi/\lambda$, and $m$ is the mass of KRb.  We use microwaves at $\sim$2.2~GHz to couple the $|0,0\rangle$ and $|1,-1\rangle$ states, which form the $|\downarrow\rangle$ and $|\uparrow\rangle$ two-level system.  The degeneracy of the $N=1$ rotational states is broken due to the interaction between the nuclear quadrupole moments and the rotation of the molecules~\cite{hyperfine}, and in a $54.59$~mT magnetic field, the $|1,0\rangle$ and $|1,1\rangle$ states are higher than the $|1,-1\rangle$ state by 270~kHz and 70~kHz, respectively (Fig.~\ref{fig:scheme}b)~\cite{noteaboutnuclearspinstates}. The quantization axis is set by the magnetic field, which is at 45 degrees with respect to the $x$ and $y$ lattice directions (Fig.~\ref{fig:scheme}c).  The polarizations of the lattice beams are chosen such that the tensor AC polarizabilities of the $|0,0\rangle$ and $|1,-1\rangle$ states are very similar~\cite{brian}, so that we create a spin-state-independent lattice trap (more information can be found in the Supplementary Material).

The energy scale for our spin-1/2 quantum XY system is characterized by $J_\perp/2 \times V_{dd}(\mathbf{r}_i-\mathbf{r}_j)$. For our rotational states, $|d_{\downarrow\uparrow}|=0.98\times0.57/\sqrt3$ Debye and $|J_\perp/(2h)|=52$~Hz.  Here, the additional factor of 0.98 in the transition dipole matrix element comes from the estimated $2\%$ admixture of another hyperfine state~\cite{hyperfine}. Each molecule in the lattice will experience an interaction energy with contributions from all other molecules, where each contribution depends on the separation of the two molecules in the lattice and the angle $\Theta$.  Figure~\ref{fig:scheme}c shows the geometrical factors for nearby sites relative to a central molecule (green) for our experimental conditions.

\begin{figure}
\includegraphics[width=16cm]{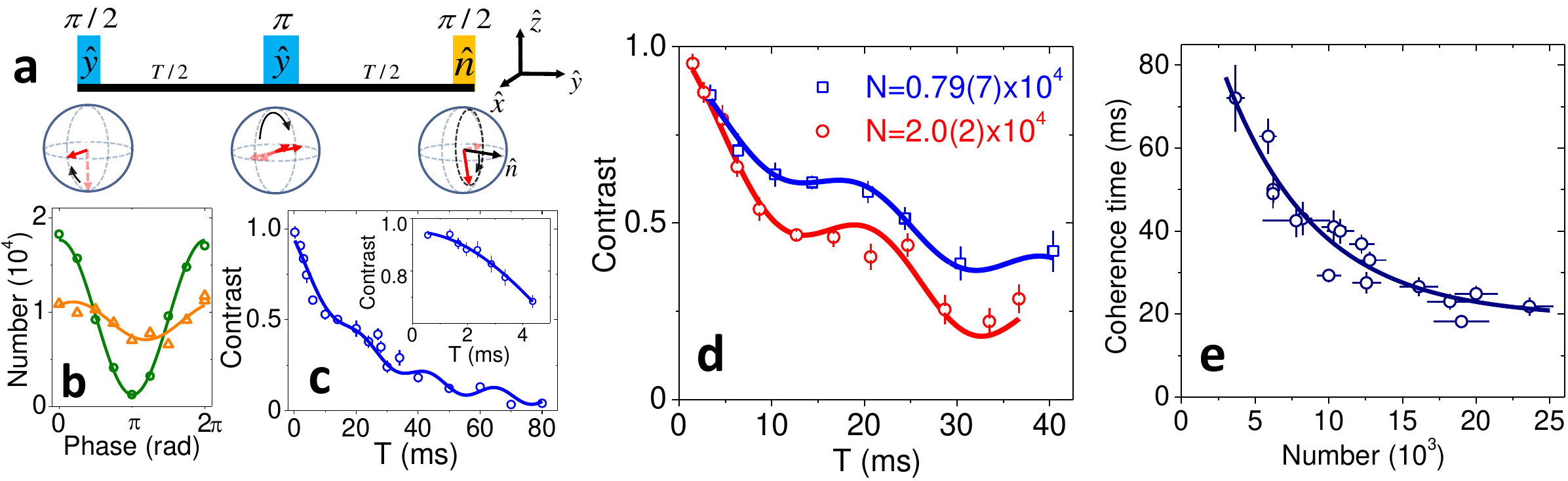}
\caption{\label{fig:density}
\textbf{Coherent spin dynamics of polar molecules.}
\textbf{~a}, A ($\pi/2)_y$ pulse initializes the molecules in a coherent superposition of rotational states.  A spin-echo pulse sequence is used to correct for effects arising from single-particle inhomogeneities across the sample, such as residual light shifts.  \textbf{b}, The phase of the final $\pi/2$ pulse is scanned (corresponding to rotations around a variable axis $\hat{n}$) to obtain a Ramsey fringe. Two fringes are shown, corresponding to the short and intermediate time scales. \textbf{c}, The contrast of the Ramsey fringe is measured as a function of interrogation time. Since the molecules' spin states are initially all in phase, at very short times, $T < 2h/J_{\perp}$, the contrast decay curve should be quadratic~\cite{hazzard:far-from-equilibrium_2013}, as shown in the inset.
\textbf{d}, The contrast of the Ramsey fringe versus interrogation time is shown for two different filling factors, characterized by the initial molecular number. In addition to the density-dependent decay, we observe clear oscillations, which arise from spin-exchange interactions between neighboring molecules.
\textbf{e}, The spin coherence time decreases for increasing molecule number. The solid line shows a fit to $C + A/N$, where $C$ and $A$ are constants.
}
\end{figure}

We employ coherent microwave spectroscopy to initiate and probe spin dynamics in our system. Figure~\ref{fig:density}a shows a basic spin-echo pulse sequence and its Bloch sphere representation. Starting with the molecules prepared in the $|\downarrow\rangle$ state, the first $(\pi/2)_y$ pulse creates a superposition state $\frac{1}{\sqrt{2}}(|\downarrow\rangle+|\uparrow\rangle)$. Because the AC polarizabilities of the two states are nearly identical, we can address the entire sample with a relatively weak microwave drive (with a Rabi frequency of 2.7~kHz), which helps to avoid off-resonant coupling to other states. For this microwave power, we obtain a fidelity of greater than $99\%$ for $\pi$ pulses.  A residual differential AC Stark shift gives rise to dephasing due to the variation in the light intensity across the sample; however, this single-particle effect can be removed using a spin-echo pulse.
After a free evolution time $T/2$, we apply a $(\pi)_y$ echo pulse, which flips the spins and thus reverses the direction of single-particle precession. The spins rephase after another free evolution time $T/2$, at which time we probe the coherence by applying a $\pi/2$ pulse with a phase offset relative to the initial pulse. We measure the number of molecules left in the $|\downarrow\rangle$ state as a function of this offset phase, which yields a Ramsey fringe (Fig.~\ref{fig:density}b).

With the single-particle dephasing effectively removed, the contrast of the Ramsey fringe as a function of $T$ yields information on spin interactions in the system~\cite{martin:Sr_2013}. We note that the spin-echo pulse has no impact on the dipolar spin-exchange interactions described by Eqn.~1. The most striking feature evident in the measured contrast curves (Fig.~\ref{fig:density}c,d) is the oscillations on top of an overall decay.  We attribute both the contrast decay and the oscillations to dipolar interactions.  Imperfect lattice filling and many-body interactions each give a spread of interaction energies, which results in dephasing and a decaying contrast in the Ramsey measurement.
Fig.~\ref{fig:scheme}c illustrates the different interaction energies coming from $V_{dd}$, which can be positive or negative.
For low lattice fillings, the interaction energy spectrum can have a strong contribution from the largest magnitude nearest-neighbor interaction. Oscillations in the contrast can then result from the beating of this particular frequency with the contribution from
molecules that experience negligible interaction shifts.  In principle, there could be several different oscillation frequencies owing to the differing geometrical factors in the lattice; however, we do not resolve multiple frequencies in the experiment.

Since interaction effects depend on the density, we investigate spin coherence for different lattice filling factors.  To reduce the density of molecules without changing the distribution, we hold the molecules in the lattice for a few seconds while inducing single-particle losses with an additional strong optical beam that enhances the rate of off-resonant light scattering~\cite{chotia:long-lived_2012}.  We fit the measured time dependence of the Ramsey contrast to an empirical function $Ae^{-T/\tau}+B\cos^2(\pi f T)$ to extract a coherence time $\tau$ and an oscillation frequency $f$.  As shown in Fig.~\ref{fig:density}d, the coherence time $\tau$ depends on the number of molecules, or filling fraction, but the oscillation frequency $f$ is essentially unchanged for our accessible range of densities.

We observe oscillation frequencies in the range $48 \pm 2$~Hz for molecule numbers in the lattice that vary by threefold. The fact that this frequency is consistent with the largest nearest neighbor interaction energy of $J_\perp/2=52$~Hz supports the conclusion that the contrast oscillations come from nearest neighbor dipole-dipole interactions. Because this frequency is determined by the lattice geometry and the dipole matrix element, it does not depend on the lattice filling factor. We also confirm that the oscillation frequency does not depend on the lattice depth from 20 to 50 $E_r$. For the coherence time, we observe a strong dependence on the filling factor (Fig.~\ref{fig:density}e).  Density dependence is a classic signature of interaction effects, and we conclude that the coherence time in the deep lattice is limited by dipole-dipole interactions.  For higher filling factors, the increasing probability that molecules have multiple neighbors means that more spin-exchange frequencies will contribute to the signal, which leads to faster dephasing.

\begin{figure}
\includegraphics[width=16cm]{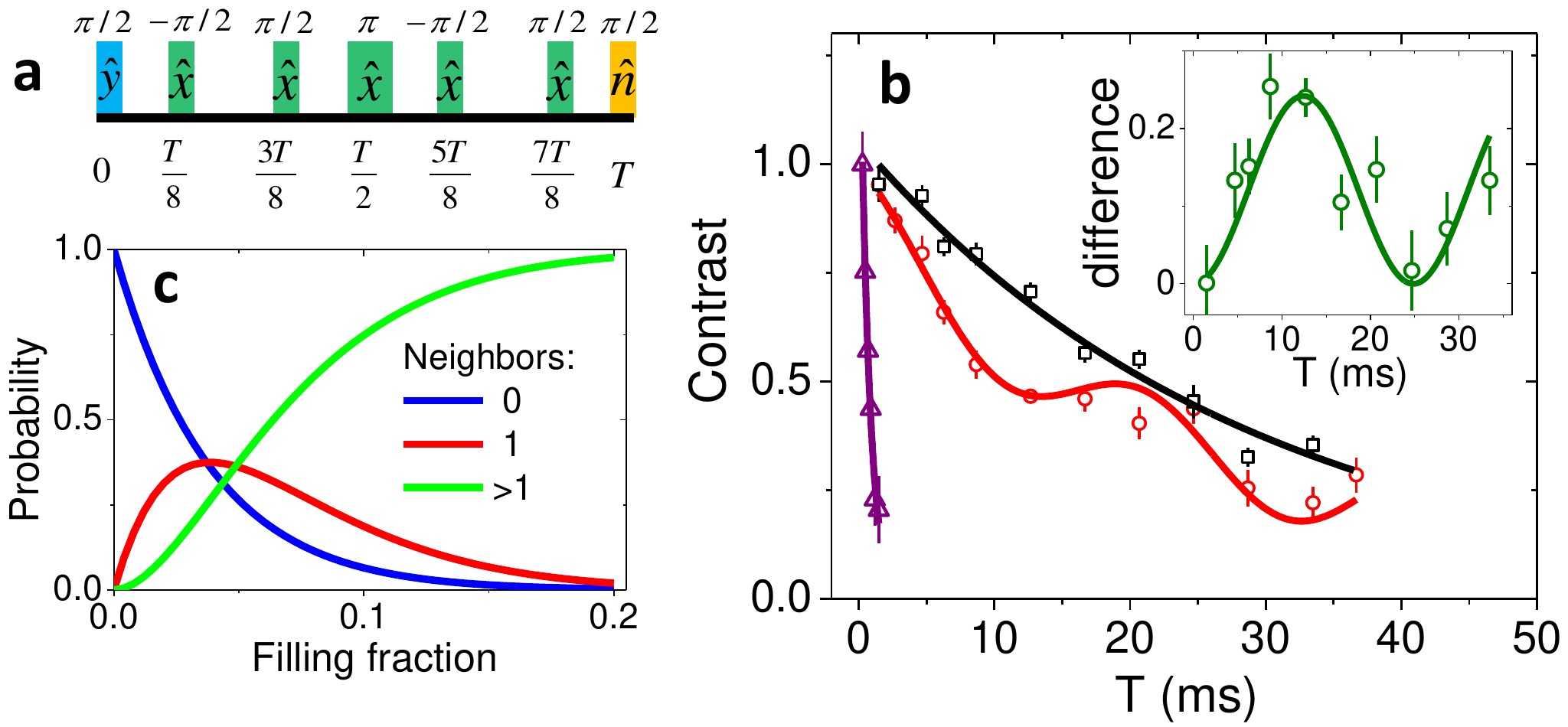}
\caption{\label{fig:decoupling}
\textbf{Multi-pulse sequence and decoupling of pair-wise dipolar interactions.}
\textbf{~a}, The multi-pulse sequence is designed to suppress both single-particle dephasing and the effect of pair-wise dipole-dipole interactions.
\textbf{b}, The contrast decay is displayed as a function of time under three different pulse sequences. Without a spin-echo pulse, single-particle inhomogeneities result in a Ramsey coherence time of about 1~ms (triangles). The spin-echo pulse effectively removes the single-particle dephasing, so that spin-exchange interactions play the dominant role in the contrast decay (circles). The multi-pulse sequence suppresses the contrast oscillations and slightly improves the coherence time (squares).  \textbf{Inset} The difference in contrast between the the multi-pulse sequence and the spin-echo case shows clear oscillations. \textbf{c}, The probability for a particular molecule to have zero, one, or more than one neighbors (within the cube shown in Fig.~\ref{fig:scheme}c) is plotted as a function of a uniform lattice filling factor.
}
\end{figure}

Multi-pulse sequences, as well as single spin-echo pulses, are examples of dynamical decoupling, which is widely used in NMR~\cite{waugh:approach_1968} and quantum information processing~\cite{du,bollinger,maurer:room-temperature_2012} to remove dephasing and extend coherence times. Although a spin-echo pulse cannot mitigate the contrast decay that arises from dipole-dipole interactions, a multi-pulse sequence can. In particular, the pulse sequence~\cite{waugh:approach_1968} shown in Fig.~\ref{fig:decoupling}a, is designed to remove dephasing due to two-particle dipolar interactions (Supplementary Material).  Analogous to how a spin-echo pulse works, this pulse sequence flips the eigenstates of the dipolar interaction Hamiltonian (Eqn.~1) for two isolated particles to allow for subsequent rephasing.

Figure~\ref{fig:decoupling}b summarizes the Ramsey contrast decay for three different pulse sequences. With a simple two-pulse Ramsey sequence (with no spin-echo pulse), the coherence time of the system is very short, with the fringe contrast decaying within 1~ms (triangles in Fig.~\ref{fig:decoupling}b). With the addition of a single spin-echo pulse, the single-particle dephasing time can be extended to $\sim$80~ms (measured for our lowest molecular density). However, this coherence time is reduced dramatically with increasing molecule number in the lattice, and we observe oscillations in the contrast signal (circles in Fig.~\ref{fig:decoupling}b). When we apply the multi-pulse sequence, the oscillations in the contrast are suppressed, and the data fit well to a simple exponential decay with a coherence time slightly longer than that of the spin-echo case (squares in Fig.~\ref{fig:decoupling}b). The differences in the measured contrast oscillations and decay for the usual spin-echo and multi-pulse sequence highlight the spin-exchange dynamics driven by pair-wise dipolar interactions (Fig.~\ref{fig:decoupling}b inset).

To understand the dynamics of this spin system, a key ingredient is the filling fraction of molecules in the 3D lattice, since the Ramsey contrast decay depends sensitively on the molecular density (cf.~Fig.~\ref{fig:density}e).
Figure~\ref{fig:decoupling}c shows the probability for a particular molecule to have zero, one, or more than one neighbors, where we consider the 26 neighboring sites shown in Fig.~\ref{fig:scheme}c and a uniform lattice filling.  We attribute the contrast oscillation to nearly isolated pairs of molecules, which can be substantial near a filling fraction of $4 \%$.
On the other hand, interactions of multiple molecules will contribute to the spin decoherence, and we note that for filling fractions greater than 5\%, the probability to have exactly one neighbor is surpassed by the probability to have two or more neighbors.

\begin{figure}\label{fig:quantum}
\includegraphics[height=10cm]{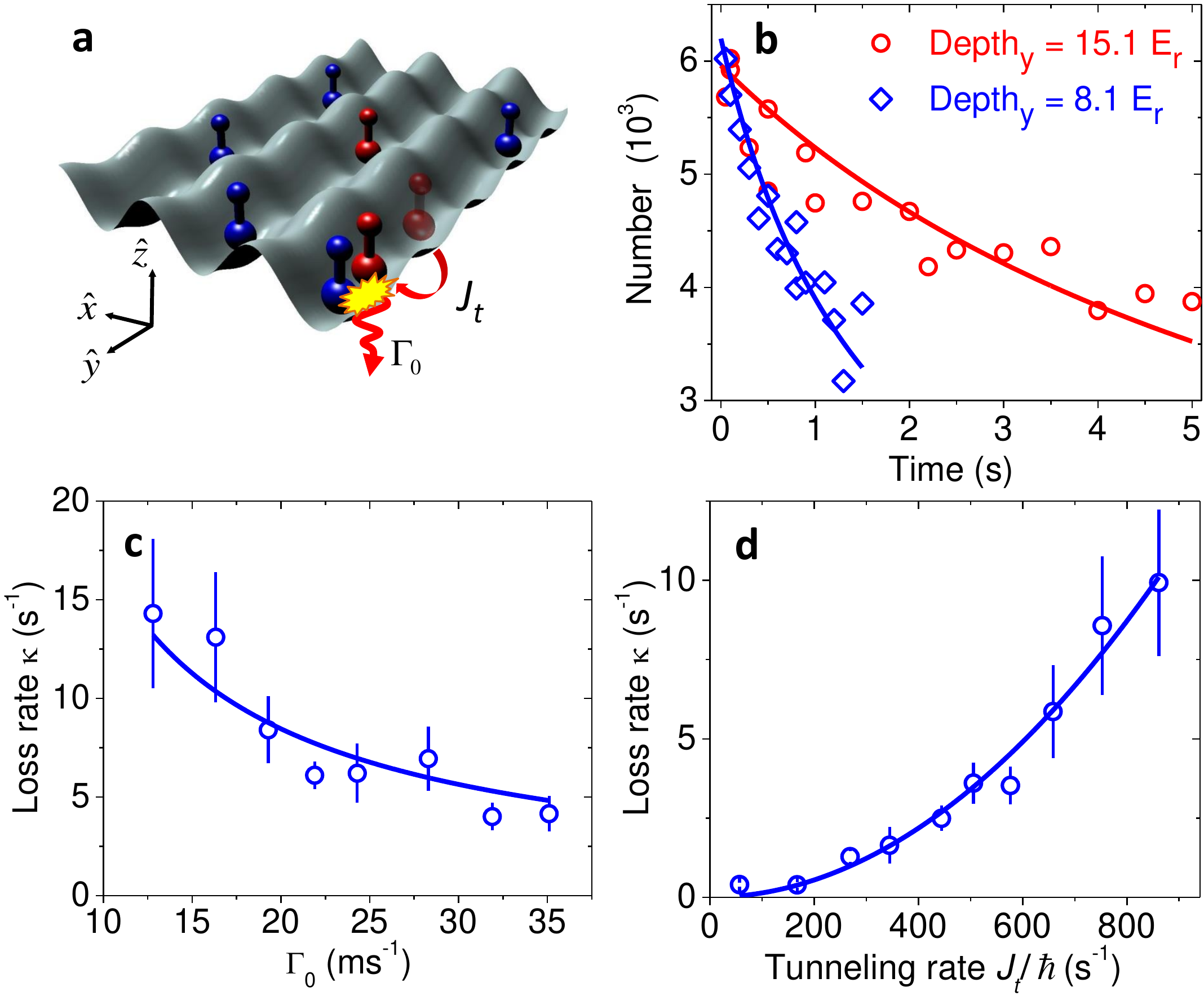}
\caption{\label{fig:zeno}
\textbf{Quantum Zeno effect for polar molecules in a 3D lattice.}
\textbf{~a},~The lattice depths along $x$ and $z$ are kept at $40$~$E_r$, while the lattice depth along $y$ is reduced to allow tunneling along the $y$ direction at a rate $J_t/\hbar$. Once two molecules in different spin states tunnel to the same site, they are lost due to chemical reactions at a rate $\Gamma_0$.
\textbf{b},~Number loss of $|\downarrow \rangle$ state molecules versus time is shown for lattice depths along $y$ of $8.1$~$E_r$ and $15.1$~$E_r$.
\textbf{c},~The number loss rate $\kappa$ versus $\Gamma_0$ fits to a $1/\Gamma_0$ dependence, which is consistent with the quantum Zeno effect.
\textbf{d},~The number loss rate $\kappa$ versus $J_t$ fits to a $(J_t)^2$ dependence, as predicted from the quantum Zeno effect.
}
\end{figure}

To provide an independent determination of the filling fraction, we have measured tunneling-induced loss at reduced lattice depths. Molecules are initially prepared in the $|\downarrow\rangle$ state in a $40 \ E_r$ lattice. For our fermionic molecules, the chemical reaction rate is much larger between molecules in distinguishable internal states~\cite{silke2010}.  Moreover, Pauli blocking strongly suppresses molecules in the same spin state from tunneling into the same lattice site. Therefore, we create a 50/50 incoherent spin mixture of $|\downarrow\rangle$ and $|\uparrow\rangle$ by applying a $\pi/2$ pulse and waiting 50~ms.   We then quickly (within 1~ms) lower the lattice depth along only a single direction ($y$, as shown in Fig.~\ref{fig:zeno}a) to allow tunneling and loss due to on-site chemical reactions~\cite{silke2010,ni2010,miranda2011}.
We then measure the remaining number of molecules in the $|\downarrow\rangle$ state as a function of the holding time.
Figure~\ref{fig:zeno}b shows example loss curves for two different lattice depths along $y$.

In our system, the on-site loss rate $\Gamma_0$ is proportional to the chemical reaction rate between the $|0,0\rangle$ and $|1,-1\rangle$ molecules~\cite{miranda2011}:
\begin{equation}
\Gamma_0=\beta\int|W(x,y,z)|^4 \, \mathrm{d}x \, \mathrm{d}y \, \mathrm{d}z,
\end{equation}
where $\beta=1.0(1)\times10^{-12}$ cm$^{-3}$ is the two-body loss coefficient~\cite{silke2010} and $W(x,y,z)$ is the ground-band Wannier function.  We can modify $\Gamma_0$ by changing the lattice depth;
however, for our measurements, the system always remains in the strongly interacting regime in which $\Gamma_0 \gg J_t/\hbar$, where $J_t$ is the tunneling amplitude.  This is the regime of the quantum Zeno effect~\cite{misra:zeno_1988,itano:quantum_1990,Streed:2006,rempe2008}, where increasing $\Gamma_0$ actually decreases the effective two-body loss rate between neighboring molecules, which is given by~\cite{rempe2008}
\begin{equation}\label{eqn:gammaeff}
\Gamma_{\text{eff}}=\frac{2(J_t/\hbar)^2}{\Gamma_0}.
\end{equation}
The number of $|\downarrow \rangle$ molecules, $N_{\downarrow}(t)$, can then be described with a two-body loss equation
\begin{equation}\label{eqn:numberloss}
\frac{dN_{\downarrow}(t)}{dt}=-\frac{\kappa}{N_{\downarrow,0}} N_{\downarrow}(t)^2,
\end{equation}
where $N_{\downarrow,0}$ is the initial number of $|\downarrow \rangle$ molecules and the loss rate coefficient is given by~\cite{baur:two-body_2010} \footnote{ We note that in Ref.
\cite{baur:two-body_2010} there is a factor of 2 missing on the right hand side of Eq. (13).} $\kappa=4q\Gamma_{\text{eff}} g_{\downarrow\uparrow}^{(2)}n_{\downarrow,0}$.
Here, $2n_{\downarrow,0} = n_0$ is the initial filling fraction in the lattice, $q=2$ is the number of nearest neighbor sites in our 1D tunneling geometry, and
$g_{\downarrow \uparrow}^{(2)}$ is the correlation function of different spin states for nearest neighboring sites $i$ and $j$: $g_{\downarrow\uparrow}^{(2)}=
\langle \hat{n}_i \hat{n}_j-4\vec{S}_i\cdot \vec{S}_j\rangle/\langle \hat{n}_i\rangle^2$, with $\hat{n}_i$ the number operator at site $i$ and $\vec{S}_i$ the spin $1/2$ vector operator.  In our case, we assume that initially the molecules
are randomly distributed in the $|\downarrow\rangle$ and $|\uparrow\rangle$ states, so that $g_{\downarrow\uparrow}^{(2)}=1$. Since the redistribution of molecules due to losses and tunneling can modify $g_{\downarrow\uparrow}^{(2)}$, we fit the data to the solution of Eqn.~\ref{eqn:numberloss} for short times, where the number has changed by less than $50\%$.

We verify the quantum Zeno effect by measuring the dependence of the loss rate $\kappa$ on $\Gamma_0$ and $J_t$. To study the dependence on $\Gamma_0$, we set the lattice depth along $y$ to be $5.4(4)$~$E_r$, which fixes $J_t$, and then increase the lattice depths along the $x$ and $z$ directions. This compresses the wave function $W(x,y,z)$ in each lattice site, and thus increases $\Gamma_0$.  As expected for the quantum Zeno regime, the measured $\kappa$ decreases as $\Gamma_0$ increases, and the data are consistent with $\kappa \propto 1/\Gamma_0$ (Fig.~\ref{fig:zeno}c). To study the dependence on $J_t$, we vary the lattice depth along $y$, while simultaneously adjusting the $x$ and $z$ lattice depths to keep $\Gamma_0$ fixed. As shown in Fig.~\ref{fig:zeno}d, the measured $\kappa$ exhibits a quadratic dependence on $J_t$ as predicted by Eqn.~\ref{eqn:gammaeff}. For these loss rate measurements, all parameters are known except the initial filling fraction $n_0$. From measurements of the loss rate at several lattice depths, we find $n_0$ to be $9(1)\%$ for $2\times10^4$ molecules. This estimate is consistent with calculations of the Ramsey fringe contrast decay using a cluster expansion (see Methods Summary).

Although it is desirable to increase the lattice filling in order to explore interesting phases such as quantum magnetism or exotic superfluidity, we have seen that the modest filling factors achieved for our experiment already enable the observation of dipolar interaction effects in a 3D lattice spin model.  Adding an external electric field would increase the variety of spin models that can be realized with this system. Our measurements suggest the intriguing possibility to study emergent phenomena expected in spin models in which disorder plays a crucial role, such as glassy phases~\cite{randeria:low-frequency_1985}, interaction-driven many-body localization~\cite{basko:many-body_2006,pal:many-body_2010,alvarez:nmr_2010}, and even energy transport in biomolecules~\cite{saikin:photononics_bio_2013}.

\section*{Methods summary}
\textbf{Preparation of molecules in a 3D optical lattice} We begin with about $1\times10^5$ $^{87}$Rb atoms and $2.5\times10^5$ $^{40}$K atoms in a far-off resonance dipole trap at 1064~nm. The trap frequencies are $25$~Hz radially and $185$~Hz axially for Rb, where the axial direction is along $\hat z$. The Rb gas forms a Bose-Einstein Condensate (BEC) with $T/T_c \approx 0.5$, while the K Fermi gas is at $T/T_F \approx 0.5$, where $T_c$ is the transition temperature for BEC and $T_F$ is the Fermi temperature. We smoothly ramp on a 3D lattice over $100$~ms to a final depth of $40$~$E_r$ ($16$ and $7$ recoil energies for Rb and K atoms, respectively). The $x$ and $y$ lattice beams have a waist of $200$~$\mu$m and the $z$ beam has a waist of $250$~$\mu$m. The lattice depth is calibrated with parametric heating of the molecular gas~\cite{brian}, and has an estimated uncertainty of $5\%$.  After turning on the lattice, we lower the intensity of the dipole trap to zero in $50$~ms, and then ramp a magnetic field from $54.89$~mT to $54.59$~mT in $1$~ms to create weakly bound KRb Feshbach molecules. We then use two-photon stimulated Raman adiabatic passage (STIRAP) to transfer the Feshbach molecules to the rovibrational ground state. The unpaired Rb and K atoms are removed using resonant light scattering.  After molecules are created in the lattice, we can perform band-mapping measurements by quickly turning off the lattice in $1$~ms. We find that the fraction of molecules in higher bands is consistent with zero within our detection limit of $5\%$. To measure the number of ground-state molecules in the lattice, we reverse the STIRAP process to recreate Feshbach molecules, and then take an absorption image using light resonant with the K cycling transition.

\textbf{Theoretical modeling of the spin dynamics}  Theoretical modeling of the spin dynamics observed with Ramsey spectroscopy shows similar oscillations and coherence times as our measurements, and the comparison can be used to estimate a filling factor of $5 - 10\%$ for $2\times10^4$ molecules. Although exactly treating the many-body dynamics is intractable, at sufficiently small filling a ``cluster expansion'' that separates $N$ molecules into $N/g$ clusters of $g$ molecules and solves exactly the spin dynamics within these clusters~\cite{witzel:quantum_2005,maze:electron_2008} can be quite accurate (we use $g$ up to 10).  For example, in the extremely dilute limit, most molecules sit far from all other molecules, with only a few occupying adjacent lattice sites. The isolated molecules have static dipolar interactions and only clusters with two (or more) particles evolve in time.

\vspace{2.5cm}

\textbf{Acknowledgments} We thank B. Zhu, M. Foss-Feig, and M. Lukin for stimulating discussions.  We acknowledge funding for this work from NIST, NSF, AFOSR-ARO (MURI), ARO, DOE, and ARO-DARPA-OLE. S.A.M. is supported by an NDSEG Graduate Fellowship. B.G. and K.R.A.H. are National Research Council postdoctoral fellows.

\textbf{Correspondence} Correspondence should be addressed to Deborah Jin (~jin@jilau1.colorado.edu) and Jun Ye~(ye@jila.colorado.edu).

\bibliographystyle{plain}

\renewcommand{\theequation}{S\arabic{equation}}
\renewcommand{\thefigure}{S\arabic{figure}}

\renewcommand{\thefootnote}{\fnsymbol{footnote}}

\renewcommand{\theequation}{S\arabic{equation}}
\renewcommand{\thefigure}{S\arabic{figure}}

\newpage

\begin{center}
\textbf{\Large Supplementary Information}
\end{center}

\noindent {\large \textbf{\textsc{1) Differential light shift in a 3D lattice}}}

Molecules have complex internal structure; hence, there are a number of different approaches to finding a magic trap that matches the polarizabilities of two different internal states. The polarizability of molecules is anisotropic, so tuning the angle between the quantization axis and the polarization of the light field can change the polarizabilities~\cite{brian}. For a 3D lattice, there are three different polarization vectors. The lattice geometry in our experiment is shown in Fig.~1c in the main text. We choose the $x$ and $y$ lattice beams to have their polarizations along the horizontal plane, at an angle of $\pm 45$ degrees relative to the magnetic field, respectively. The $z$ lattice polarization is the same as $x$ lattice. Following our previous work~\cite{brian}, the energy shifts for the $|1,0\rangle$, $|1,-1\rangle$, and $|1,1\rangle$ states are determined by finding the eigenvalues of the Hamiltonian
\setcounter{equation}{0}
\begin{equation}
H=-\alpha(45^\circ)I_x-\alpha(-45^\circ)I_y-\alpha(45^\circ)I_z+\mathrm{diag}(\epsilon_1,\epsilon_2,\epsilon_3),
\end{equation}
where $I_x$, $I_y$, and $I_z$ are the intensities of lattice beams along the $x$, $y$, and $z$ directions, $\alpha$ is the polarizability matrix defined in reference\cite{brian}, and $\epsilon_1$, $\epsilon_2$, and $\epsilon_3$ are the energies for $|0,0\rangle$, $|1,-1\rangle$, and $|1,1\rangle$, respectively.

Figure~\ref{fig:magic} shows the differential light shift (with respect to the $|0,0\rangle$ state) of the $|1,0\rangle$, $|1,-1\rangle$, and $|1,1\rangle$ states as a function of the lattice depth. The $|1,-1\rangle$ state has the smallest intensity dependence, which corresponds to the minimal inhomogeneity due to light shifts. The inset shows an expanded plot for the $|1,-1\rangle$ state. The red points are the experimentally measured transition frequencies for different lattice depths, which agree well with theory. When the lattice depth is about $40 E_r$ in each direction, the differential light shift is zero. We measure the transition frequency between the $|0,0\rangle$ and $|1,-1\rangle$ states in a $40 \, E_r$ lattice to be $2.227 783 38(8)$~GHz, which agrees with the measured frequency of $2.227 783 35(4)$~GHz in the absence of any optical potentials.
At this lattice depth, the slope for the differential light shift is 120 $\text{Hz}/E_r$, and the variation of the light shift across the entire sample is less than $500$ Hz.

\setcounter{figure}{0}
\begin{figure}
\includegraphics [height=6cm]{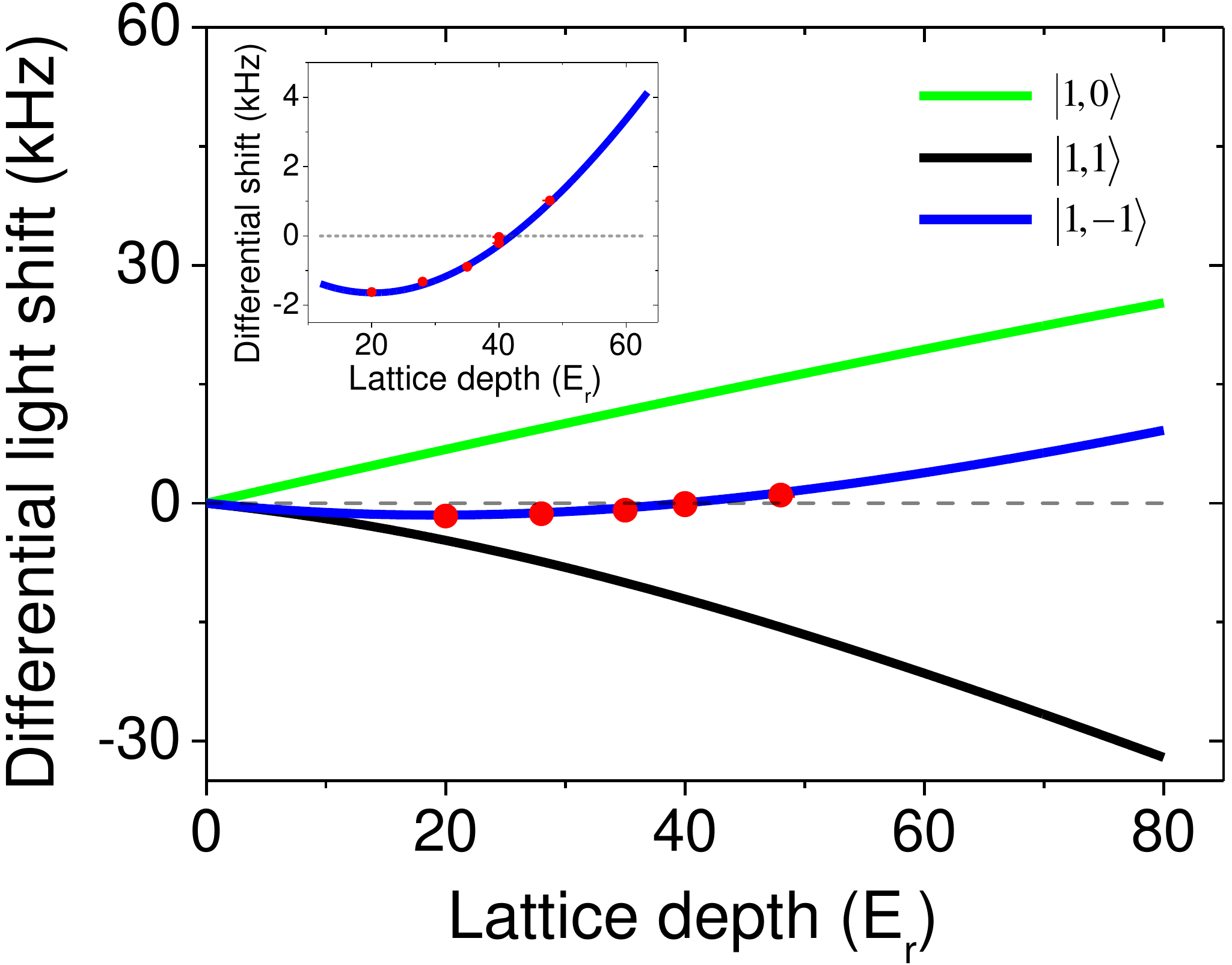}
\caption{\label{fig:magic}
\textbf{The differential light shift in a 3D lattice.} The differential light shift (with respect to the $|0,0\rangle$ state) is shown as a function of the lattice depth. Green, black, and blue lines show calculated light shifts for $|1,0\rangle$, $|1,1\rangle$, and $|1,-1\rangle$, respectively.  Red points are experimental data for the $|1,-1\rangle$ state. Inset: An expanded view for the $|1,-1\rangle$ state is displayed. The differential light shift between $|1,-1\rangle$ and $|0,0\rangle$ is zero when the lattice depth is about 40 $E_r$ in each beam.}
\end{figure}

\noindent {\large \textbf{\textsc{2) Multi-pulse sequence}}}

It is straightforward to understand how the multi-pulse sequence works for two particles. Two molecules are initially prepared in the $|\downarrow\downarrow \rangle$ state.  The %
first $(\pi/2)_y$ pulse transfers them to
\begin{equation}                                                                                                                               \frac{1}{\sqrt{2}}(|\downarrow\rangle+|\uparrow\rangle)\otimes\frac{1}{\sqrt{2}}(|\downarrow\rangle+|\uparrow\rangle)=\frac{1}{2}(|\downarrow\downarrow\rangle+|\uparrow\uparrow\rangle+|\downarrow\uparrow\rangle+|\uparrow\downarrow\rangle).   \end{equation}
Because of the spin exchange term, $|\downarrow\uparrow\rangle$ and $|\uparrow\downarrow\rangle$ are not eigenstates of the Hamiltonian expressed in Eqn. 1 of the main text. However, the three triplet states $|\downarrow\downarrow\rangle$, $|\uparrow\uparrow\rangle$, and $\frac{1}{\sqrt{2}}(|\downarrow\uparrow\rangle+|\uparrow\downarrow\rangle)$ are eigenstates of the Hamiltonian, with eigenenergies 0, 0, and $J_\perp/2$. During the first free evolution time $T/8$,  $|\downarrow\downarrow\rangle$ and $|\uparrow\uparrow\rangle$ accumulate no phase, while $\frac{1}{\sqrt{2}}(|\downarrow\uparrow\rangle+|\uparrow\downarrow\rangle)$ accumulates a phase $e^{-i(J_\perp/\hbar)T/16}$.  At this point the state is entangled.  Then we apply a $(-\pi/2)_x$ pulse, which transforms the $|\downarrow\uparrow\rangle+|\uparrow\downarrow\rangle$ state to                                              $|\downarrow\downarrow\rangle+|\uparrow\uparrow\rangle$, and the $|\downarrow\downarrow\rangle+|\uparrow\uparrow\rangle$ state to $|\downarrow\uparrow\rangle+|\uparrow\downarrow\rangle$. Another way to view this is that the $(-\pi/2)_{x}$ pulse swaps the phases between the  $|\downarrow\uparrow\rangle+|\uparrow\downarrow\rangle$ and $|\downarrow\downarrow\rangle+|\uparrow\uparrow\rangle$ states.  After another $T/4$ evolution time, the $(\pi/2)_x$ pulse swaps the phases again. This state then freely evolves for another $T/8$, after which both $|\downarrow\uparrow\rangle+|\uparrow\downarrow\rangle$ and $|\downarrow\downarrow\rangle+|\uparrow\uparrow\rangle$ have accumulated the same phase $e^{-i(J_\perp/\hbar)T/8}$, and so the state is no longer entangled. In this way the dephasing due to dipole-dipole interactions is canceled. The center $(\pi)_x$ pulse and another pair of $(-\pi/2)_x$ and $(\pi/2)_x$ pulses are necessary for removing the single-particle inhomogeneity in addition to rephasing the dipole-dipole interactions. The effects of dipole-dipole interactions beyond that of isolated pairs of molecules are not removed by this particular multi-pulse sequence.

\begin{figure}
\includegraphics [height=6cm]{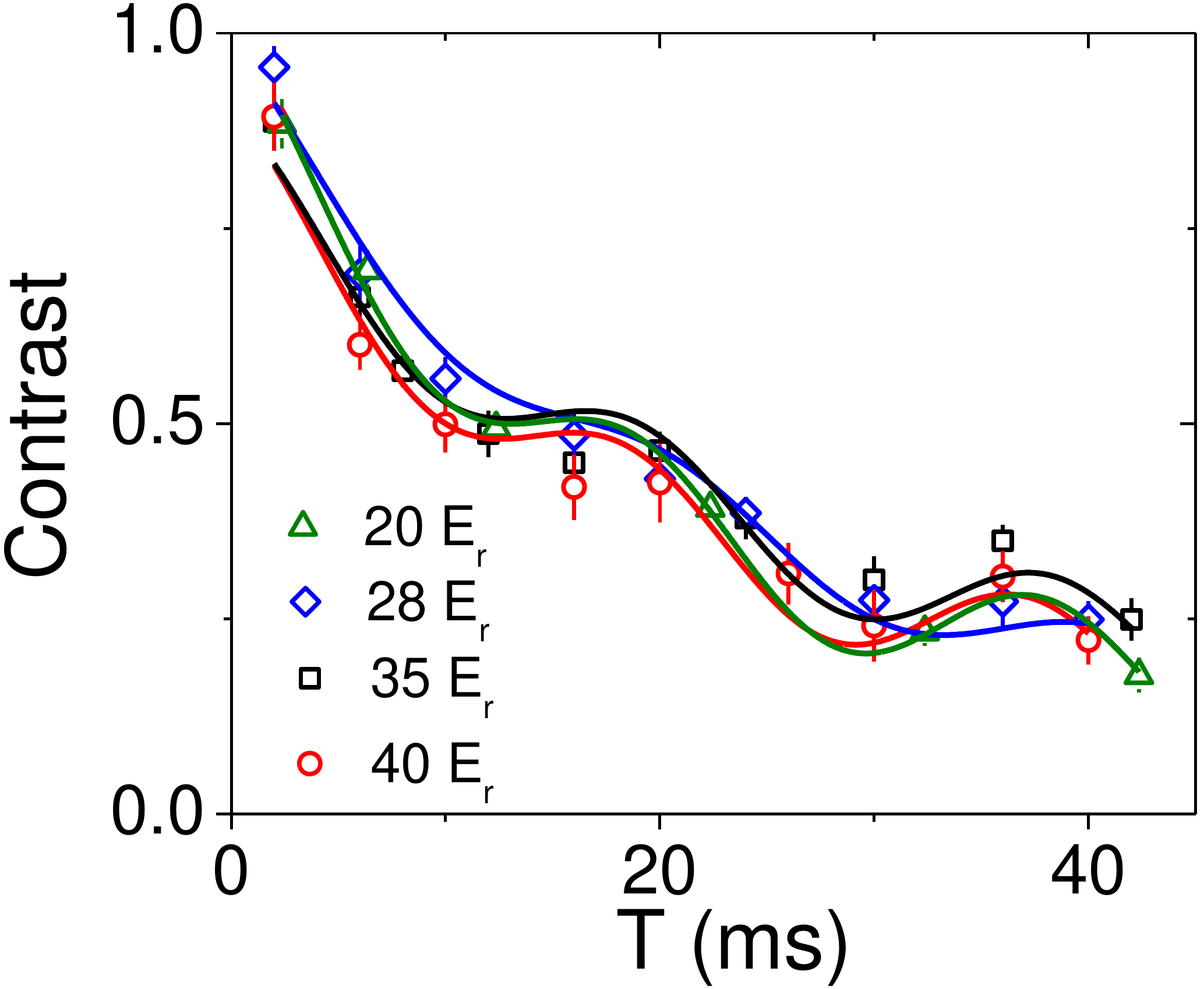}
\caption{\label{fig:intensitydependence}
\textbf{The contrast oscillation at different lattice depth.} Four different Ramsey contrast measurements have been taken for lattice depths of 20, 28, 35, and 40 $E_r$. The oscillation frequency in the contrast stays invariant with respect to the lattice depth within our measurement precision. }
\end{figure}

\vspace{4mm}
\noindent {\large \textbf{\textsc{3) Independence of the contrast oscillation on lattice depth}}}
To demonstrate that the oscillation frequency in the Ramsey fringe contrast does not depend on the lattice depth, we show in Fig.~\ref{fig:intensitydependence} four different Ramsey contrast measurements taken when the lattice depth is varied from 20 to 40 $E_r$.

\bibliography{bibdipole-3}

\end{document}